\documentclass{article}
\usepackage[lmargin=3cm,rmargin=3cm,bmargin=3cm,tmargin=3cm]{geometry}
\usepackage[section]{placeins} 
\usepackage{amsmath,amsfonts,amssymb,amsthm}
\usepackage{natbib}
\usepackage{color}
\usepackage{todonotes}
\usepackage{bm} 
\usepackage{tikz}
\usepackage{booktabs}
\usepackage{hyperref}

\usetikzlibrary{matrix, chains, arrows}
\newcommand{\s}{\sigma}
\newcommand{\n}{\mathbf n}
\newcommand{\id}{\iota}

\newcommand{\signedperm}{\overline{\pi}}

\newtheorem{example}{Example}[section]

\tikzstyle{bpnot}= [circle,inner sep=0pt, minimum size=5pt]
\tikzstyle{n}= [bpnot,fill=black]
\tikzstyle{g} = [draw, >=stealth', thick]
\tikzstyle{gh} = [g,->]
\tikzstyle{gt} = [g,<-]
\newcommand{\adjgraphvertices}[5]{
  \node (source) at #1 {};
  \node[bpnot] (last) at (source) {};
  \foreach \chr in {#2} {
  \edef\last{}
    \foreach \gene / \s in \chr {
      \ifthenelse{\equal{\s}{+}}{\def\ea{t}\def\eb{h}}{\def\ea{h}\def\eb{t}}
      \node[n,label=#3:$\last\gene_{\ea}$] (#4_\ea\gene) at (last) {};
      \xdef\last{\gene_{\eb}}
      \node[n,right=of last] (last) {};
      \ifnum #5>0 
        \draw[g\eb,shorten >= 1pt, shorten <= 1pt] (#4_\ea\gene) -- node[above,black] {$\mathbf{\s\gene}$} (last);
      \fi
      \node[bpnot] (#4_\eb\gene) at (last) {};
    }
    \node[n,label=#3:$\last$] at (last) {};
  }
}

\newcommand{\signedpermutation}{
\pgfmatharray{\sp}{0}\let\n\pgfmathresult
\pgfmathparse{360.0/\n}\let\segment\pgfmathresult
\pgfmathparse{\segment/2}\let\shift\pgfmathresult
\def\radius{1.5cm}
\def\labelrad{1.7cm}
\def\regionboundaryin{1.4cm}
\def\regionboundaryout{1.6cm}
\scalebox{1.3}{\begin{tikzpicture}
\foreach \x in {1,2,...,\n}
{
  \draw[thick] (360-\x*\segment+90:\regionboundaryin)
             --(360-\x*\segment+90:\regionboundaryout);
  \pgfmatharray{\sp}{\x}\let\tmp\pgfmathresult 
  \pgfmathparse{int(\tmp)}\let\tmp\pgfmathresult 
  \node at (360-\x*\segment+90+\shift:\labelrad) {\tmp};
  \pgfmathparse{360-(\x-1)*\segment+90}\let\alpha\pgfmathresult;
  \pgfmathparse{360-(\x-1)*\segment+90-\segment}\let\beta\pgfmathresult;
  \pgfmathgreater{\tmp}{0}\let\decision\pgfmathresult
  \ifnum \decision=1
    \draw[color=black,very thick,>=latex,->] (\alpha:\radius) arc (\alpha:\beta:\radius);
  \else
    \draw[color=black,very thick,>=latex,->] (\beta:\radius) arc (\beta:\alpha:\radius);
  \fi
};
\end{tikzpicture}}}
\newcommand{\unsignedpermutation}{
\pgfmatharray{\sp}{0}\let\n\pgfmathresult
\pgfmathparse{360.0/\n}\let\segment\pgfmathresult
\pgfmathparse{\segment/2}\let\shift\pgfmathresult
\def\radius{1.5cm}
\def\labelrad{1.7cm}
\def\regionboundaryin{1.4cm}
\def\regionboundaryout{1.6cm}
\scalebox{1.3}{\begin{tikzpicture}
\foreach \x in {1,2,...,\n}
{
  \draw[thick] (360-\x*\segment+90:\regionboundaryin)
             --(360-\x*\segment+90:\regionboundaryout);
  \pgfmatharray{\sp}{\x}\let\tmp\pgfmathresult 
  \pgfmathparse{int(\tmp)}\let\tmp\pgfmathresult 
  \node at (360-\x*\segment+90+\shift:\labelrad) {\tmp};
  \pgfmathparse{360-(\x-1)*\segment+90}\let\alpha\pgfmathresult;
  \pgfmathparse{360-(\x-1)*\segment+90-\segment}\let\beta\pgfmathresult;
  \pgfmathgreater{\tmp}{0}\let\decision\pgfmathresult
  \ifnum \decision=1
    \draw[color=black,very thick] (\alpha:\radius) arc (\alpha:\beta:\radius);
  \else
    \draw[color=gray,very thick] (\beta:\radius) arc (\beta:\alpha:\radius);
  \fi
};
\end{tikzpicture}}}

\title{Position and content paradigms in genome rearrangements: \\ the wild and crazy world of permutations in genomics.}
\author{Sangeeta Bhatia, Pedro Feij\~ao, Andrew R. Francis} 
\begin{document}
\maketitle
\begin{abstract}
Modellers of large scale genome rearrangement events, in which segments of DNA are inverted, moved, swapped, or even inserted or deleted, have found a natural syntax in the language of permutations.  Despite this, there has been a wide range of modelling choices, assumptions and interpretations that make navigating the literature a significant challenge. Indeed, even authors of papers that use permutations to model genome rearrangement can struggle to interpret each others' work, because of subtle differences in basic assumptions that are often deeply ingrained (and consequently sometimes not even mentioned).
In this paper, we describe the different ways in which permutations have been used to model genomes and genome rearrangement events, presenting some features and limitations of each approach, and show how the various models are related. This paper will help researchers navigate the landscape of genome rearrangement models, and make it easier for authors to present clear and consistent models.
\end{abstract}

\section{Introduction}
\label{sec:intro}
While the order of genes along a chromosome has been used to extract phylogenetic information as far back as 1938~\citep{dobzhansky1938inversions}, it wasn't until almost half a century later that the problem of determining distance between gene arrangements was formalized~\citep{watterson1982chromosome}.  Right from the beginning, the language and notation of permutations were used, and have been part of the field ever since.  

Different arrangements are thought to arise due to various large scale changes in the genome, by which we mean changes affecting more than just a single nucleotide.  Examples of such changes include the reversal of a segment of the genome (called an \emph{inversion} or \emph{reversal}), the movement of a segment to a different location on the genome (\emph{transposition}),  events that split a chromosome in two or join two chromosomes into one single chromosome (\emph{fission} and \emph{fusion}) and the exchange of segments between different chromosomes (\emph{translocation}). The rearrangement distance is defined to be the minimal number of rearrangement events between a
pair of genomes. The set of rearrangement events through which genomes are hypothesised to have changed could consist of a single type of event such as inversion, or a combination of events such as inversion, translocation, fusion and fission.

Representing the contrasting genome arrangements on paper is best done with some form of permutation notation.  Rearrangement events, then, are performed on the representation of the arrangement. The action of the rearrangement events depends of course on the choice of notation. 

There are two main philosophies of representing genomes as permutations and this has generated some confusion in the literature, as it is frequently not clear which paradigm is being used.  In this paper we call these paradigms  ``position'' and ``content'', and while each generates permutations that can look similar, they have significantly different meanings, and rearrangement events are consequently implemented differently in each.  In fact there is a clear relationship between them, and in this article we describe them both, their philosophies, their limitations, and the relationship between them.

The structure of the paper is as follows.  We begin in Section~\ref{sec:notation} with some standard notation for permutations that will be used throughout.  The following three sections form the body of the paper and for each paradigm describe respectively how genomes are represented (Section~\ref{sec:genomemaps}); how rearrangements are represented (Section~\ref{sec:operator}); and how the paradigms are related (Section~\ref{sub:links}).

\section{Permutation notation}\label{sec:notation}

\subsection{Basic definitions}
A permutation on a set $X$ is a bijection from $X$ to itself. Without loss of generality, the set $X$ is usually considered to be the set of integers $\mathbf{n}=\{1,2, \dots n\}$. A permutation $\pi$ can be written in ``two-line notation'' by listing all the elements of the domain and their images under $\pi$.
\begin{example}\label{ex:perm}
Let $\pi$ be the permutation of the set $\n=\{1, 2, 3, 4, 5\}$ that maps $1\mapsto 4$, $2\mapsto 3$, $3\mapsto 5$, $4\mapsto 1$, and $5\mapsto 2$.  Its representation in two-line notation is given below:
\[
\pi = \begin{pmatrix}
1 & 2 & 3 & 4 & 5\\
4 & 3 & 5 & 1 & 2
\end{pmatrix}.
\]
\end{example}
As the first line of the two-line notation is always the same, it is common to use only the bottom row to represent a permutation, for instance writing the permutation $\pi$ in Example~\ref{ex:perm} as $\pi = [4, 3, 5, 1, 2]$.  We will use square brackets and commas to denote this representation of a permutation as a list of images of 1 up to $n$.

Every permutation can also be expressed as a product of disjoint cycles. We write a cycle in a permutation as $(i_1\ i_2\ \dots\ i_k)$, meaning $i_1$ is mapped to $i_2$, $i_2$ to $i_3$, etc, with $i_k$ mapped back to $i_1$. 
The cyclic decomposition of the permutation $\pi$ in Example~\ref{ex:perm} is $(1\ 4)(2\ 3\ 5)$.

The \emph{identity permutation} is the permutation $\iota=[1, 2, \dots , n]$. Every permutation $\pi$ has a unique \emph{inverse}, denoted by $\pi^{-1}$, that satisfies $\pi^{-1}\pi= \pi \pi^{-1} = \iota$.
In cycle notation, the inverse can be obtained by simply writing each cycle down in reverse order.  For two-line notation, writing the inverse of $\pi$ requires mapping each number to its position in the second line of the two-line notation for $\pi$.

The set of all permutations on a set of size $n$ forms an algebraic structure called the \emph{symmetric group} usually denoted as $S_n$. The multiplication operation in the symmetric group is the familiar composition of functions. 

It is also possible (and common) to represent genomes in ways that incorporate an orientation on the regions.  The set of ``signed'' permutations is used for this, and this set forms a group called the \emph{hyperoctahedral group}.  However there are some other subtleties associated with signed permutations and we defer discussing those to Section~\ref{s:orientation}.

\subsection{Actions}

The first complexity in representing permutations in cycle notation comes when one chooses whether the permutation acts on the \emph{left} or on the \emph{right}.  While it is common in much of mathematics to think of functions acting on the \emph{left}, so that we might write $f(x)$ for $f$ acting on an element $x$ of the domain, it is also common in the study of group actions to write functions acting on the \emph{right}, so that we might write $(x)f$ or simply $xf$ for $f$ acting on $x$.  

For those unfamiliar with this latter usage, it might begin to make more sense when considering function composition: when acting on the left, $f g$ means we ``do $g$ first'', while when acting on the right it means ``do $f$ first'', which accords with the way we read (in ``left-to-right'' languages).  The other intuitive feature of actions on the right for permutations is that the reading of an individual cycle goes from left to right, (so that $(2\ 3\ 5)$ means $2$ maps to $3$, $3$ maps to $5$, and $5$ maps to $2$), and this then agrees with the order in which a composition of cycles is read.   Note the contrast with the action on the left: in this case a composition of cycles is read from the right, but within each cycle we read left to right.

Both these conventions appear in the literature on genome rearrangements, and so we will describe both, taking care to make it clear at each point which convention is in play.  Note that the map described above for writing a permutation as a product of disjoint cycles is valid regardless of which convention is used, because a) within each cycle we always read left to right, and b) disjoint cycles commute (the action is the same regardless of the order in which they are performed).

The choice of whether our permutations act on the right or on the left will become important when we discuss how to implement rearrangements such as inversions on the genome, using cycle notation.  This will be discussed in Section~\ref{sec:operator}.

\section{Genome representation}\label{sec:genomemaps}

We use the words `genes' and `regions' interchangeably to mean `conserved regions' or what are sometimes called `synteny blocks' or `locally colinear blocks' in genome rearrangement literature.  Essentially these denote regions of DNA that are present in all of the genomes under study, and whose different arrangements are of interest.  
The word `genome' is used to indicate the set of chromosomes involved (and sometimes there is just one).

\subsection{The position paradigm -- Genomes as maps between positions and regions} \label{subsec:classic}

The first paradigm we discuss is the explicit use of \emph{position} to describe a chromosome.  This approach has been used widely in the literature beginning with the seminal papers by \citet{sankoff1992gene,bafna1993genome} to recent developments such as~\citet{egrinagy2013group}.  The description of a chromosome here consists of saying which {position} a gene or region is in, or equivalently, what region is in each position.  These two versions may seem hard to distinguish but they reflect two viewpoints that lead to opposite (or inverse) notation, and so we will spend some time expanding on this theme. 

The starting point, adopted widely, is to use the two-line notation described above to put a genome into permutation language.  This involves either writing the position numbers along the top of an array, and the region numbers along the bottom, or vice versa:
\[
\pi=\begin{pmatrix}
1 & 2 & \cdots & n \\
\pi_1 & \pi_2 & \cdots & \pi_n 
\end{pmatrix}\ :\quad\text{either}\quad 
\begin{pmatrix}
 \text{positions}\\
 \text{regions}
\end{pmatrix}\quad\text{or}\quad
\begin{pmatrix}
 \text{regions}\\
 \text{positions}
\end{pmatrix}
\]
If the positions are labelled in sequence around each chromosome, then the bottom row of the ``positions to regions'' version is just the labels of the regions read along the genome.

This notation, for both representations, effectively treats a chromosome as a map $\pi$ between the set of positions $\n=\{1,2,\dots,n\}$ and the set of regions.

It is common to denote the set of regions also by the integers $1,\dots n$, which, while natural, can lead to confusion because the map $\pi$ then looks like a bijection on the set $\n$.  It is important to realize that, in this genome representation, this map is not a bijection on the set $\n$, but a one-to-one correspondence between two different sets: one positions, one regions. 
(Multi-chromosomal genomes are usually modelled as a collection of permutations where each permutation encodes a chromosome (e.g.~\citet{kececioglu1995mice,hannenhalli1995transforming})). 
It is also clear that as permutations, the two representations produce inverses of each other: one maps a position to the region that is in that position, and the other maps a region back to its position.

When we write a permutation (genome arrangement) in cycle form, it will also be different depending on which representation we use.

For example, the genome 
\begin{center}
\resizebox{.2\textwidth}{!}{
\def\sp{{6,2,4,1,3,6,5}}\unsignedpermutation
}
\end{center}
has two-line and cycle notation forms
\[
\text{pos}\to\text{reg}: \begin{pmatrix}
1 & 2 & 3 & 4 & 5 & 6\\
2 & 4 & 1 & 3 & 6 & 5
\end{pmatrix}\ =\ (1\ 2\ 4\ 3)(5\ 6)
\]
(the 4-cycle should be read ``position 1 has region 2, and position 2 has region 4, and position 4 has region 3'' and so on), and
\[
\text{reg}\to\text{pos}: \begin{pmatrix}
1 & 2 & 3 & 4 & 5 & 6\\
3 & 1 & 4 & 2 & 6 & 5
\end{pmatrix}\ =\ (1\ 3\ 4\ 2)(5\ 6)
\]
(which should be read ``region 1 is in position 3, and region 3 is in position 4'', etc). Note that $[(1\ 2\ 4\ 3)(5\ 6)]^{-1}=(1\ 3\ 4\ 2)(5\ 6)$.

Although \citet{watterson1982chromosome} do not explicitly write the gene arrangements in cycle form, their description of a genome consists of mapping regions to positions. Similarly \citet{egrinagy2013group} also write the arrangements with the set of regions constituting the domain and the positions as the co-domain. This choice is more pragmatic from the point of view of software implementation. We will elaborate on this point in Section~\ref{sec:operator}. 

In the rest of the literature, genome arrangement as a map from positions to regions is more commonly used \citep{Bafna1998,sankoff1992gene,hannenhalli1999transforming}. In particular, modeling linear genomes as an ordered list of genes is a natural choice. 

Note that this paradigm of genome representations requires referring to an absolute position for each region.  This must be chosen \emph{a priori} in order to write down the genome.  This choice must then be taken into account when considering the distance between two genomes, as mentioned in Section~\ref{s:rearr.position}.

\subsubsection{Incorporating orientation} 
\label{s:orientation}
To incorporate the orientation of DNA into the models, various approaches have been used. The most common is to use \emph{signed permutations}, where the sign of a region represents its orientation. 
A signed permutation is a permutation of the set $\{ -n, \dots,-1,1,\dots, n\}$, where it is common to assume that it satisfies $\pi_{-i} = -\pi_i$ (e.g., \citet{Moulton2012,Labarre2011}).  

For instance, the genome
\begin{center}
\resizebox{.2\textwidth}{!}{
\def\sp{{6,2,-4,1,-3,6,5}}\signedpermutation
}
\end{center}
has two-line form
\[
\text{pos}\to\text{reg}: \left(\begin{array}{*{12}{r}} 
-6 & -5 & -4 & -3 & -2 &-1 & 1 & 2 & 3 & 4 & 5 & 6\\
-5 & -6 & 3 & -1 & 4 & -2 & 2 & {-4} & 1 & {-3} & 6 & 5
\end{array} \right) 
\]
which in cycle notation form is $(1\ 2\ {-4}\ 3)({-1}\ {-2} \ 4 \ {-3})(5\ 6)({-5}\ {-6})$. 
Note that the cycles come in pairs, and given the constraint that $\pi_{-i} = -\pi_i$, one of each is redundant.  Consequently, this permutation may be abbreviated to $(1\ 2\ {-4}\ 3)(5\ 6)$.
The representation in the ``reg $\to$ pos'' form is the inverse of this permutation.

Another way dealing of with orientation is to translate the problem into the realm of unoriented or unsigned permutations. This was done for example by \citet{hannenhalli1995transforming}. Let $\signedperm: \n \rightarrow \{-n,\dots,-1,1,\dots n\}$ be a one-to-one function where $| \signedperm(i) | \neq |\signedperm(j)|$ for $i \neq j$, that is, each position $i$ points to a different gene $\signedperm(i)$, and therefore $\overline{\pi}$ represents a signed chromosome. $\signedperm$ can be associated with a permutation $\pi$ on the set $2\n= \{1, 2,\dots, 2n\}$ where $\pi$ is related to $\signedperm$ by Equation~\eqref{equn:signedtoun}.

\begin{equation}
\label{equn:signedtoun}
\pi := 
\begin{cases}
\pi(2i-1)=2\signedperm(i)-1 , \quad \pi(2i)=2\signedperm(i) & \mbox{ if } \signedperm(i) > 0 \\
\pi(2i-1)=2\signedperm(i) , \quad \pi(2i)=2\signedperm(i) - 1  & \mbox{ if } \signedperm(i) < 0.
\end{cases}
\end{equation}
For instance, the signed chromosome above would be represented by the permutation 

\[
\pi =  \left(\begin{array}{*{12}{r}} 
1 & 2 & 3 & 4 & 5 & 6 & 7 & 8 & 9 & 10 & 11 & 12\\
3 & 4 & 8 & 7 & 1 & 2 & 6 & 5 & 12 & 11 & 10 & 9
\end{array} \right). 
\]

\subsection{The content paradigm -- Genomes as maps from regions to regions} \label{subsec:alternate}

The second dominant modeling approach is to view a chromosome as a map from the set of regions to itself. 

\subsubsection{Genomes as cycles}\label{s:genomes.cycles}
For example, in what is probably the first formal discussion of reversal distance problem, \citet{watterson1982chromosome} described a chromosome as a set of neighboring gene loci (regions). 
This view of a chromosome was first formalised in terms of permutations by \citet{meidanis2000alternative} who use the cycle structure of a permutation to capture a circular chromosome. The notion of $i$ being ``mapped to'' $j$ in a cycle is interpreted as region $i$ being followed by region $j$ on the chromosome. Therefore, a circular chromosome is represented by the single cycle $\pi = (\pi_1 \ \pi_2 \ \dots \ \pi_n)$.
Multi-chromosomal genomes are then represented by many disjoint cycles, one for each chromosome. 
Meidanis and Dias also used signed permutations to model orientation, representing 
each chromosome by \emph{two} cycles, one for the direct
orientation and other for the reverse complement, with the property that
$\pi_{-i} = -\pi_{i}^{-1}$ (note this is different from the convention for signed permutations described in Section~\ref{s:orientation}).  

For instance, the circular signed chromosome from
Section~\ref{s:orientation} is modelled by 
$$\pi = (2\ {-4}\ 1 \ {-3}\ 6\ 5)(-5\ {-6}\ 3\ {-1}\ 4\ {-2}).$$  
This representation has been used to study some rearrangement distances, such as fission, fusion and transposition~\citep{Dias2001}, block-interchange~\citep{Huang2010}, and 2-break operations~\citep{feijao2013extending}.

Although multi-chromosomal genomes are easier to write down, when compared to the ``positional'' permutations, it is clear that this formalism is more suitable for dealing with circular genomes than linear genomes, since chromosomes are permutation cycles.  
\subsubsection{Genomes as adjacencies}\label{s:genomes.adj}
In order to develop a permutation model that allows for multi-chromosomal genomes, with both linear and circular chromosomes, it was necessary to use an alternative formulation, where the focus is shifted from the ordering of the genes to the connections between genes. This model became one of the most common ways of representing genomes in the combinatorial community (e.g. \citet{bergeron2006unifying,Tannier2009}). A \emph{gene} is defined as an oriented section of the DNA, and its two ends are called its \emph{extremities}. To represent a genome, considered as an arrangement of oriented genes, it is sufficient to note which extremities are adjacent on the genome. An (unordered) pair of extremities that are adjacent is referred to as an \emph{adjacency}. An extremity that is not adjacent to any other is the end point of a linear chromosome and is called a \emph{telomere}. 

This model can easily uniquely describe a multi-chromosomal genome by its set of adjacencies and telomeres, even if the genome contains both linear and circular chromosomes. The \emph{genome graph} is a graph where the vertices are the adjacencies and telomeres of a genome (sets of one or two gene extremities), and for each gene there is a directed edge from the vertex with the tail of the gene to the vertex with the head of the gene.  A \emph{path} in this graph corresponds to a linear chromosome, and a \emph{cycle} to a circular chromosome. It is easy to recover the gene order and orientation in this graph by traversing the components, labelling the edges with the gene label and a plus or minus sign, depending on the direction of the traversal and edge orientation.
In the rest of this paper, we refer to this genome representation as the \emph{adjacency list} model.

An algebraic formulation of the adjacency list model was presented by \citet{feijao2013extending} and independently by \citet{bhatia2014algebraic}. 
In the algebraic framework, the gene extremities may be represented by the set of signed integers $\{-n, \dots, -1, 1, \dots n\}$~\citep{feijao2013extending} or by mapping the set of $2n$ gene extremeties into the set $\{1, 2, \dots, 2n\}$~\citep{bhatia2014algebraic}. An adjacency is represented as a $2$-cycle and a genome is expressed as a product of disjoint $2$-cycles ($1$-cycles implicitly represent the telomeres and are usually omitted, since they represent fixed points in the permutation). 

\begin{figure}
\begin{center}
\begin{tikzpicture}[node distance=6em]
  \adjgraphvertices{(0,0)}{{{1/+},{3/-},{2/+},{4/+}}}{above}{a}{1}  
  \foreach \cyc in {(1),(2 \ 6),(5 \ 3),(4 \ 7),(8)} {
  \node[bpnot,label=below:${\cyc}$] at (source) {};
  \node[bpnot,right=of source] (source) {};
}
\end{tikzpicture}
\end{center}
\caption{The genome graph of a genome with one linear chromosome containing genes numbered 1, 2, 3 and 4. The vertex set of the graph is the set of adjacencies and extremities, $\{1_t,1_h3_h,3_t2_t,2_h4_t,4_h\}$. Directed edges are drawn from the tail to the head of the same gene. 
To express this chromosome as a permutation $\pi$, the set of extremities $\{1_t,1_h,2_t,2_h, \hdots, 4_h\}$ is mapped into $ \{1,2, \hdots, 8 \}$ (\citet{bhatia2014algebraic}) or into $\{+1,-1,\dots,+4,-4\}$~(\citet{feijao2013extending}). $1_h$ is connected to $3_h$ which is captured by the 2-cycle $(2 \ 6)$ in the permutation encoding. The other 2-cycles can be similarly interpreted. The above genome is thus encoded as the permutation $(2 \ 6)(5 \ 3)(4 \ 7)$ or  $(-1 \ {-3})(3\ 2)(-2\ 4)$.}
\label{fig:adjlist}
\end{figure}
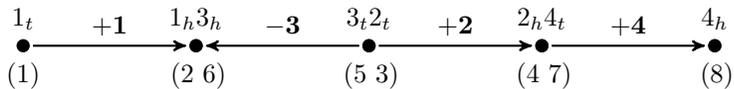

Figure~\ref{fig:adjlist} shows a genome graph consisting of a linear chromosome, together with its permutation representation as an adjacency list.

\section{Genome rearrangement events} \label{sec:operator}

As we have seen in Section~\ref{sec:genomemaps}, a genome may be viewed as a map between sets of positions and regions, or from a set of regions to itself.  In this section we describe how a rearrangement event can be also modelled as a map that acts on a genome, using the language of permutations. The rearrangement event map should therefore be \emph{composable} with the genome, meaning that as functions the appropriate matchings of domains and codomains occurs. 

We begin with the position paradigm.

\subsection{Rearrangements in the position paradigm}\label{s:rearr.position}

A rearrangement in this paradigm is an action on the \emph{positions} of the genome.  So, for instance, an inversion swapping adjacent positions 2 and 3 swaps those two positions regardless of what regions are in those positions.  This is really the logic behind this paradigm, as an operation that acted on \emph{regions} 2 and 3 would affect those two regions regardless of where they physically were located on the genome (this is a feature of rearrangements in the ``content'' paradigm, described in Section~\ref{s:rearr.regions}).  

Thus, in this view of modeling genomes, a rearrangement event is a permutation on the set of positions.  

The observation that the position paradigm involves a map between positions and regions, while a rearrangement is a map from positions to positions, imposes constraints on how they may be composed.  These constraints, in turn, depend on whether we wish to write our group actions (rearrangement events) acting on the \emph{left} or on the \emph{right}.  The permissible compositions are shown in Table~\ref{tab:permitted.actions}.

\begin{table}[ht]\centering
\caption{Correct ways to compose a rearrangement such as inversion (``inv'') with a genome (permutation).  The constraints on the action arise because of the needs for codomain of the first function to match the domain of the second function.  For example, after applying a genome permutation representation that takes positions to regions, one cannot then act by a rearrangement function that takes positions to positions. }
\label{tab:permitted.actions}
\begin{tabular}{@{}llll@{}} 
\toprule
\begin{minipage}{2cm}Permutation\\ representation:\end{minipage} & Composition: & Legal action: & Used?\\
\midrule
reg $\to$ pos & ([reg]) (reg $\to$ pos) $\cdot$ inv & Only on the right & Yes\\
pos $\to$ reg & \phantom{([reg]) }(reg $\leftarrow$ pos) $\cdot$ inv ([pos]) & Only on the left & Yes\\
pos $\to$ reg & ([pos]) inv $\cdot$ (pos $\to$ reg) & Only on the right & No\\
reg $\to$ pos & \phantom{([pos]) }inv $\cdot$ (pos $\leftarrow$ reg) ([reg]) & Only on the left & No\\
\bottomrule
\end{tabular}
\end{table}

\begin{figure}[ht]  
\begin{center}
\begin{tikzpicture}

\draw (0,0) node (newg) {\resizebox{.2\textwidth}{!}{
\includegraphics{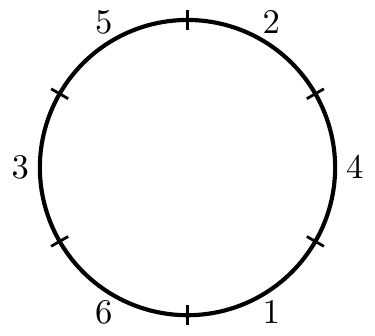}}};
\node[below=0.05 cm of newg] {new genome};
\node[above= 1.5cm of newg] (newp){\begin{minipage}{1.5cm}\centering new\\ positions\end{minipage}};
\node[above= of newp] (p) {positions};
\node[above= 1.5cm of p] (g) {\resizebox{.2\textwidth}{!}{
\includegraphics{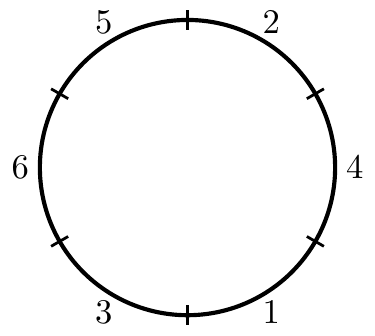}
}};
\node[above=0.05 cm of g] {genome};
\node[left=1.5cm of p] (r1) {regions};
\node[right=1.5cm of p] (r2) {regions};
\node[left=1.5cm of newp] (r3) {regions};
\node[right=1.5cm of newp] (r4) {regions};

\draw[>=latex,->] (r1)-- node[above](pi1){} (p);
\node[draw, left=0.05cm of r1, align=center] {$(1\ 3\ 4\ 2)(5\ 6)$};
\draw[>=latex,->] (r1)-- node[below](pi){$\pi$} (p);
\draw[>=latex,->] (r3)--node[above]{$\pi \cdot \rho$} node[below](pi3){}(newp);
\node[draw, below left=0.05cm of r3, align=center] {$(1\ 3\ 4\ 2)(5\ 6)\cdot(4\ 5)$ \\ $=(1\ 3\ 5\ 6\ 4\ 2)$};
\draw[>=latex,->] (p)--node[above](pi2){} (r2);
\node[draw, right=0.05cm of r2, align=center] {$(1\ 2\ 4\ 3)(5\ 6)$};
\draw[>=latex,->] (p)-- node[below]{$\pi$} (r2);
\draw[>=latex,->] (newp)--node[above]{$\pi \cdot \rho$ } node[below](pi4){}(r4);
\node[draw, below right=0.05cm of r4, align=center] {$(1\ 2\ 4\ 3)(5\ 6)\cdot(4\ 5)$ \\
$=(1\ 2\ 4\ 6\ 5\ 3)$};
\draw[>=latex,->] (p)--node[right]{$(4\ 5)$}(newp);
\draw[>=latex,->] (p)--node[left]{$\rho$}(newp);
\node[above=6mm of pi1] (pi1a){};
\node[above=6mm of pi2] (pi2a){};
\node[below=6mm of pi3] (pi3a){};
\node[below=6mm of pi4] (pi4a){};

\draw[>=latex,->] (g)..controls(pi1a).. node[left]{\begin{minipage}{2cm}
\centering \emph{right\\ composition}\end{minipage}} (pi1);
\draw[>=latex,->] (g)..controls(pi2a).. node[right]{\begin{minipage}{2cm}
\centering \emph{left\\ composition}\end{minipage}}(pi2);
\draw[>=latex,->] (pi3)..controls(pi3a)..(newg);
\draw[>=latex,->] (pi4)..controls(pi4a)..(newg);

\end{tikzpicture}
\end{center}
\caption{This figure captures the difference in the left and right actions of a rearrangement operator on a circular genome. The left side shows the representation of the genome as a map from regions to positions with a right group action (the first row of Table~\ref{tab:permitted.actions}), while the right side shows the  genome as a map from positions to regions with a left group action (the second row of the table).  The vertical arrow $\rho$ indicates the action of a rearrangement event on the positions of the genome. The effect of this position change on the maps is given in the lower horizontal arrows.}
\label{fig:l-r.actions}
\end{figure}

Another way to represent the different approaches in the first two rows of Table~\ref{tab:permitted.actions} is given in Figure~\ref{fig:l-r.actions}.  
This figure brings one further subtlety in the difference between the two representations to light.  On the left side of the figure, we have ``group actions on the right''.  That is, the group elements multiply (compose) on the right, so that $\rho_1\rho_2$ means that $\rho_1$ is done first, then $\rho_2$. It means that if we have a genome $\pi$, which is a map from regions to positions, and then we perform several rearrangements on it, first $\rho_1$, then $\rho_2$, and so on up to $\rho_n$, then we can write this:
\[
\pi\cdot \rho_1\rho_2\dots\rho_n.
\]
However, on the right hand side of the figure, in which we have ``group actions on the left'', the same sequence of inversions must be written $\rho_n\dots\rho_1$.  This means that performing this sequence of rearrangements on the genome $\pi$ involves the composition:
\[
\pi\cdot \rho_n\dots\rho_2\rho_1.
\]

Note that the action of an additional rearrangement $\rho_{n+1}$ with this representation involves inserting it into the \emph{middle} of the above expression so that the new expression is $\pi\cdot \rho_{n+1}\rho_n\dots\rho_2\rho_1$, whereas in the other representation such an additional rearrangement simply involves multiplying on the right: $\pi\cdot \rho_1\rho_2\dots\rho_n\rho_{n+1}$.
For this reason, computational implementations such as~\citet{egri2014bacterial} use the former representation as mentioned in Section~\ref{sec:genomemaps}. 

A summary of this issue for each paradigm in Table~\ref{tab:permitted.actions} is given in Table~\ref{tab:LR.actions}.

As pointed out in Section~\ref{sec:genomemaps}, ~\citet{egri2014bacterial} use the regions to positions description of a genome. The choice of the groups acting on the right thus makes this work an example of the first row of Table~\ref{tab:permitted.actions}.

On the other hand, in \citet{bafna1993genome} and the subsequent papers based on their work (e.g.~\citet{caprara1997sorting}), the genome is described as a map from positions to regions and the rearrangement operators acts on the left. For example, the inversion operator  $\rho(i,j)$ that inverts the positions from $i$ to $j$ is the permutation $[1,\dots,i-1,j,j-1,\dots, i+1, i, j+1,\dots]$. Referring to the second row in Table~\ref{tab:permitted.actions}, for any position, we first find its image under $\rho$ and then look at the region mapped to this new position under the genome $\pi$. For example, $\rho$ maps $i$ to $j$ and $\pi$ maps $j$ to $\pi(j)$. In the resulting genome, $i$ is mapped to $\pi(j)$. By considering the action of $\rho$ on the other positions, we see that $\rho(i,j)$ acts on a genome $[\pi(1),\dots,\pi(i-1),\pi(i),\pi(i+1),\dots, \pi(j), \pi(j+1),\dots]$ to produce the genome $[\pi(1),\dots,\pi(i-1),\pi(j),\pi(j-1),\dots, \pi(i+1), \pi(i), \pi(j+1),\dots]$. 

\begin{table}[ht]
\caption{Different ways to represent the genome and the action of rearrangements, with a focus on the composition of rearrangement operators.  In the (right hand) Structure column the placement of an additional rearrangement $\rho_{n+1}$ is shown in bold.  Note that when the side of the group action and the side the rearrangement acts on the genome clash (one right, one left), new rearrangements must be added into the \emph{middle} of the expression.  This is not ideal, and why we recommend use of the structures in the first and last rows, both involving representations of genomes mapping regions to positions. } \label{tab:LR.actions}
\begin{tabular}{@{}cccll@{}} 
\toprule
 Structure & \multicolumn{2}{c}{Side of } & \multicolumn{1}{c}{Structure} & Example\\ 
 \cmidrule{2-3}
& group action & rearrangement action \\
\midrule
(reg $\to$ pos) $\cdot$ inv     & Right & Right & $(x)\pi\cdot \rho_1\dots\rho_n\bm{\rho_{n+1}}$ & ($x$ is a region)\\
(reg $\leftarrow$ pos) $\cdot$ inv  & Left & Right & \phantom{$(x)$}$\pi\cdot \bm{\rho_{n+1}}\rho_n\dots\rho_1(x)$ & ($x$ is a position)\\
inv $\cdot$ (pos $\to$ reg)     & Right & Left & $(x)\rho_1\dots\rho_n\bm{\rho_{n+1}}\cdot \pi$ & ($x$ is a region)\\
inv $\cdot$ (pos $\leftarrow$ reg)  & Left & Left & \phantom{$(x)$}$\bm{\rho_{n+1}}\rho_n\dots\rho_1\cdot \pi(x)$ & ($x$ is a position)\\
\bottomrule
\end{tabular}
\end{table}

Finally, as mentioned in Section~\ref{subsec:classic}, the numbering of positions around a circular genome is arbitrary, and it is difficult to talk of two genomes as having the ``same'' labelling.  In fact, the choice of labelling will have a strong effect on the minimal distance, and needs to be taken into account.  If there are $n$ regions, then there are $n$ positions and this gives rise to $2n$ distinct labellings of positions on a circle: there are $n$ choices for the first position, and two orientations one could use.  Considering these $2n$ alternative labellings is equivalent to considering the action of the \emph{dihedral group} on the set of arrangements, as noted in~\citep{egrinagy2013group,solomon2003sorting}. Similarly \citet{watterson1982chromosome} also consider the symmetries of a circular genome in determining the minimal distance while \citet{chen1996sorting} factor in the rotational symmetry i.e., the fact that a circular chromosome does not have a fixed 12 o' clock position.

\subsection{Rearrangements in the content paradigm}\label{s:rearr.regions}
When a genome is modeled as a map from a set of regions to itself, a rearrangement event must be modeled as a similar map and may act on either left or right. The effect of left and right actions will of course be different, but both these formulations have been used in the literature.

Consider for example the model of a circular genome as a cycle~\citep{meidanis2000alternative} where $i$ is mapped to $j$ if region $i$ is adjacent to region $j$ on the genome (Section~\ref{s:genomes.cycles}). 
Suppose the genome is represented as the $k$-cycle $\pi$, and $u$ is a region on the genome.
A transposition event on $\pi$ that exchanges blocks of length $i$ and $j$ starting at region $u$, 
is exchanging the blocks from $u$ to $\pi^{i-1}(u)$ and from $\pi^{i}(u)$ to $\pi^{i+j-1}(u)$ (with $i+j<k$).  To implement this as permutation multiplication, we simply multiply the cycle $\rho=(u\ \pi^i(u)\ \pi^{i+j}(u))$ on the left of $\pi$ (thinking of the permutations as acting on the left).

For instance, for the genome representation $\pi = (1\ 2\ 3\ 4\ 5\ 6)$, the cycle $\rho=(2 \ 4\ 5)$ will transpose regions 2 and 3 with region 4 (recalling here we are acting on the left):   
\[\rho\pi=(2 \ 4\ 5)(1\ 2\ 3\ 4\ 5\ 6) = (1 \ 4 \ 2 \ 3 \ 5 \ 6).\]

In this system, the permutation $\rho$ encoding the transposition event depends on the permutation $\pi$ (the genome) it is acting on. This was commented on by \citet{meidanis2000alternative} who note that this makes the problem of determining the rearrangement distance different from the group theoretic problem of determining the word length of a group element under a (fixed) generating set. 

\subsection{Rearrangement operators on adjacency lists} \label{subsec:dcjandothers}

A popular operator acting on the adjacency list representation of a genome is the ``double cut and join'' (DCJ) operator introduced by \citet{yancopoulos2005efficient} and \citet{bergeron2006unifying}. The popularity of this operator can be attributed to its simplicity and the fact that it is able to simulate most of the common rearrangement events that have been observed in comparison of genomes. 

Recall the definition of a genome graph from Section~\ref{subsec:alternate}. A DCJ operator acts on a genome graph by choosing two adjacencies at which to ``cut'' the genome graph. The four cut extremities may now be joined in one of two possible ways, each way resulting in a different rearrangement event on the genome. Other possibilities are: to cut one adjacency and join one of the extremities with a telomere; to cut one adjacency into two telomeres; or the inverse operation of joining two telomeres.

\begin{figure}
\begin{center}
  
 \begin{tikzpicture}[node distance=5em]
    \adjgraphvertices{(0,8em)}{{{1/+},{3/+},{2/+},{4/+}}}{below}{a}{1}  
  \draw[gh] (11em,5em) -- node[right,black] {\small{$\quad \{1_h3_t,2_h4_t\} \rightarrow \{1_h2_h,3_t4_t\}$}} (11em,2em);
  \adjgraphvertices{(0,0)}{{{1/+},{2/-},{3/-},{4/+}}}{below}{a}{1}  
 \end{tikzpicture}

\end{center}
\caption{The double cut and join operator can simulate several rearrangement events. In this figure, a double cut is applied at adjacencies $1_h3_t$ and $2_h4_t$, and these extremities are joined to form $1_h2_h$ and $3_t4_t$. In the chromosome at the bottom, the segment between $1_h$ and $4_t$ is reversed.}
\label{fig:dcjaction}
\end{figure}
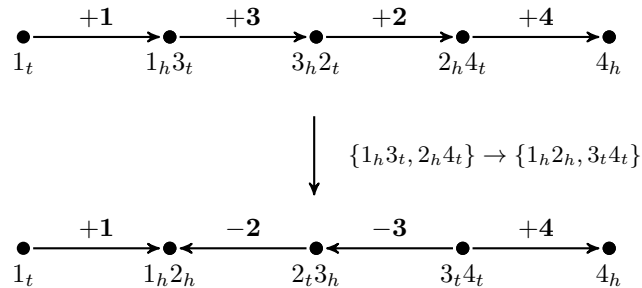

Figure~\ref{fig:dcjaction} shows a signed reversal of a section being obtained through the double cut and join operator.

\def\adjlist{$1_t$, ${1_h,3_t}$, ${3_h,2_t}$, ${2_h,4_t}$, $4_h$}

Recall from Section~\ref{subsec:alternate} that a genome can be expressed as a product of disjoint $2$-cycles. In this representation, a double cut and join operation takes the form of multiplication by two $2$-cycles. \citet{feijao2013extending} for example present it as multiplication on the left by two $2$-cycles while \citet{bhatia2014algebraic} formulate the double cut and join operation as conjugation by a $2$-cycle. For instance, consider a linear chromosome with adjacencies $(i \ j)$ and $(k \ l)$.  A DCJ that cuts these adjacencies and joins the adjacencies $(i \ k)$ and $(j \ l)$ can be obtained by conjugating the genome $\pi$ by the 2-cycle $(k \  j)$,  since 

\[(j \ k)\big[(i \ j)(k \ l)\big](j \ k) = (i \ k)(j \ l).\]

The same effect can be obtained by left or right multiplication by two 2 cycles, 
  \[(i \ l)(j \ k)\big[(i \ j)(k \ l)\big]= (i \ k)(j \ l) \ \]
and equivalently,
  \[\big[(i \ j)(k \ l)\big](j \ k)(i \ l)= (i \ k)(j \ l).\]

In fact, there is a simple relationship between these two ways of applying a DCJ operator. In general, when multiplied on the left of a genome $\pi$, a DCJ operation is of the format $(u \ v)(\pi u \  \pi v)$~\citep{feijao2013extending}. It is easy to see that $(u \ v)(\pi u \  \pi v) \pi = (u \ v)\pi(u \ v)$, which is the DCJ formulation of \citet{bhatia2014algebraic}.
The effect of the operator on the genome depends on where the regions $u$ and $v$ lie on the genome. 

\citet{feijao2013extending} generalise this operator to a $k$-break operator, where $k$ adjacencies are cut, and $k$ new ones are created from the $2k$ free extremities. As a left operator, the general form of the $k$-break operator is $(a_1 \ a_2 \ \dots \ a_k)(\pi a_k \  \pi a_{k-1} \  \dots \  \pi a_1)$, which means that a DCJ operation is a \emph{2-break}, the special case for $k=2$. Interestingly, the $k$-break operator can also be seen as a conjugation of a genome $\pi$ by a $k$-cycle:
\[(a_1 \ a_2 \ \dots \ a_k)(\pi a_k \  \pi a_{k-1} \  \dots \  \pi a_1) \pi = (a_1 \ a_2 \ \dots \ a_k)\pi(a_k \ a_{k-1} \ \dots \ a_1).\]

\section{The genome rearrangement distance problem} \label{sec:rearrdis}

Given two genomes $\pi$ and $\s$, we want to find a sequence of rearrangement operations that minimally transforms $\pi$ into $\s$. If we consider rearrangements applied on the left, we want to find $\rho_1,\rho_2, \dots, \rho_k$ such that
\[\rho_k \rho_{k-1}\cdots \rho_2 \rho_1 \pi = \s,\]
and $k$ is minimal. The \emph{rearrangement distance} between $\pi$ and $\rho$ is defined
as $d(\pi,\rho)=k$. Finding the minimal $k$ is called the \emph{rearrangement distance problem},
and finding a sequence of rearrangements transforming one genome to the other is 
commonly called the \emph{rearrangement sorting problem}.

If the rearrangement operators are invertible, then they generate a group. If starting with a single permutation (genome), say the $\id$ permutation, all (unsigned) permutations of the genes can be obtained by the repeated application of the operators under consideration, then the operators generate the symmetric group. Otherwise, the operators generate a subgroup of the symmetric group that is, a group structure that is a subset of the symmetric group.

In the position paradigm of genome representation, the problem of calculating the rearrangement distance is equivalent to finding the \emph{length of a reduced word} for a group element.  To see this, recall from above that the distance problem amounts to having two genomes, written $\pi$ and $\s$, and wanting to know the minimal number of rearrangements $\rho_i$ such that $\rho_k \rho_{k-1}\cdots \rho_2 \rho_1 \pi = \s$.  The properties of group multiplication allow us to solve this by considering $s\pi^{-1}$ (now a map from positions to positions), and factorising it in terms of the rearrangement operators.  This is precisely the problem of finding a minimal expression for a group element in terms of the generators of the group. Various researchers have noted the connection between the problems of determining rearrangement distance and finding a minimal word for a group element (e.g. see \citet{bafna1993genome,Bafna1998}).

\section{Going from one genome representation to the other} 
\label{sub:links}

In this section, we will show how to transform genomes from one representation to another. Unless otherwise noted, the compositions are acting on the left. Clearly, all results could be rewritten with right acting compositions.

\subsection{From position to content representation and back}

If $\pi$ is a genome in the regions-to-positions representation (the position paradigm in Section~\ref{subsec:classic}), then we can transform it into a unique content representation $\s$ by conjugating the $n$ cycle $(1 \ 2 \ \dots \ n)$ by $\pi$:
\[\pi(1 \ 2 \ \dots \ n)\pi^{-1}=\s,\]
where $\s$ is the genomes-as-cycles content representation from Section~\ref{s:genomes.cycles}, and where we write our permutations acting on the left (acting on the right swaps $\pi$ and $\pi^{-1}$). 
This transformation can be understood as follows.  Conjugating a $k$-cycle by an element $g$ of $S_n$ where $S_n$ is the symmetric group on a set of size $\mathbf{n}$, results in applying $g$ to every element of the cycle. That is,
\[ 
g (i_1 \  i_2 \  \dots \  i_k) g^{-1} = (g(i_1) \  g_(i_2) \  \dots  \  g(i_k)).
\]
Thus, when $\pi$ conjugates the cycle $(1\ 2\ \dots\ n)$, we obtain 
\[ \pi(1\ 2\ \dots\ n)\pi^{-1} = (\pi(1) \  \pi(2) \  \dots  \ \pi(n)) \ \]
which is exactly the ``genomes-as-cycles'' content representation.

A similar transformation can be used for signed permutations. If $\pi$ is a signed permutation in the regions-to-positions representation, satisfying $\pi_{-i} = - \pi_i$ (from Section~\ref{s:orientation}), then the signed permutation $\sigma$ in the genomes-as-cycles representation is obtained by 
\[\pi(1 \ 2 \ \dots\ \ n)(-n \ {-n}+1 \ \dots \ {-1})\pi^{-1}=\s.\]

To go from the genomes-as-cycles to the positions-to-regions representation, note that in the genomes-as-cycles representation the information about the position of each region is available relative to the other regions. Since the labeling of the positions can begin at any region, it is easy to see that multiple positions-to-regions permutations will map into the same genomes-as-cycles permutations. 

\begin{figure}
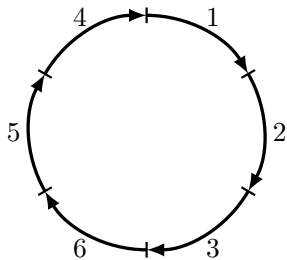

\begin{center}
\def\sp{{6,1,2,3,6,5,4}}\scalebox{.8}{\signedpermutation}
\caption{The circular genome encoded by the cycle $\sigma=(1 \ 2 \ 3 \ 6 \ 5 \ 4)$ in the content paradigm. 
}
\label{fig:circular}
\end{center}
\end{figure}

For example, consider the cycle $\s = (1 \ 2 \ 3 \ 6 \ 5 \ 4)$. Viewed as a list of
adjacent regions, it encodes the chromosome in Figure~\ref{fig:circular}. Any
of the six regions can be thought of as being at position $1$, and this assignment
then determines the position-to-regions permutation. See Figure~\ref{fig:sixpreimages} for the
permutations that correspond to $\s = (1 \ 2 \ 3 \ 6 \ 5 \  4)$. The first genome $\pi$
can be obtained from $\s$ by reading $\s$ as the bottom row of a two-line
notation (instead of as a cycle), and all the others can be obtained by
multiplying $\pi$ on the right by successive powers of the $n$-cycle $(n \  n-1 \  \dots  \  2 \ 1)$, which
corresponds to a ``rotation'' on the permutation $\pi$.  There are another $n$ labellings obtained by counting round the circle in the opposite direction (omitted here).

This multiplicity of representations arises from the symmetry inherent in a circular chromosome. Every rotation of a circular arrangement with respect to a fixed reference is equivalent. In fact this symmetry needs to be accounted for while determining rearrangement distance between a pair of circular chromosomes. This was done for example by \citet{watterson1982chromosome} and more recently by \citet{egrinagy2013group} and \citet{serdoz2016maximum}, who also consider the symmetry between a circular permutation and its mirror images. 

\def\genome{1,2,3,6,5,4}
\def\genomerad{2.0cm}
\def\innercircle{1.5cm}
\def\gap{10}

\begin{figure}
\begin{center}
  \begin{tikzpicture}
    \matrix [matrix of math nodes, ampersand replacement=\&]
   {
     \begin{pmatrix} 
       1 & 2 & 3 & 4 & 5 & 6 \\ 
       1 & 2 & 3 & 6 & 5 & 4 
     \end{pmatrix} \& 
     \begin{pmatrix} 
       1 & 2 & 3 & 4 & 5 & 6 \\ 
       4 & 1 & 2 & 3 & 6 & 5 
     \end{pmatrix} \& 
     \begin{pmatrix} 
       1 & 2 & 3 & 4 & 5 & 6 \\ 
       5 & 4 & 1 & 2 & 3 & 5 
     \end{pmatrix}  \\
     \begin{pmatrix} 
       1 & 2 & 3 & 4 & 5 & 6 \\ 
       5 & 6 & 4 & 1 & 2 & 3 
     \end{pmatrix} \& 
     \begin{pmatrix} 
       1 & 2 & 3 & 4 & 5 & 6 \\ 
       3 & 6 & 5 & 4 & 1 & 2 
     \end{pmatrix} \&  
     \begin{pmatrix} 
       1 & 2 & 3 & 4 & 5 & 6 \\ 
       2 & 3 & 6 & 5 & 4 & 1 
     \end{pmatrix} \\  
   };
\end{tikzpicture}
\end{center}
\caption{The six permutations that encode the genome in Figure~\ref{fig:circular} as a map from positions to regions, obtained by rotating the labelling of positions.  The cyclic (content paradigm) notation for the permutation encoded by these permutations is $(1 \ 2 \ 3 \ 6 \ 5 \  4)$. }
\label{fig:sixpreimages}
\end{figure}

\subsection{From adjacency lists to chromosomes}

\begin{figure}
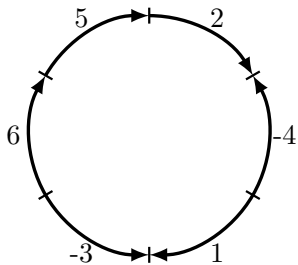

  \begin{center}
  \resizebox{.3\textwidth}{!}{
  \def\sp{{6,2,-4,1,-3,6,5}}\signedpermutation
  }
  \end{center}
  \caption{An example genome with oriented regions.}\label{fig:oriented.genome}
\end{figure}

In the content representation, we saw that there are two interpretations for the cycles of a permutations: the original algebraic formulation of \citet{meidanis2000alternative}, with cycles corresponding to chromosomes (Section~\ref{s:genomes.cycles}), and 
the adjacency list representation~\citep{feijao2013extending,bhatia2014algebraic}, with cycles as  adjacencies (Section~\ref{s:genomes.adj}). Although seemingly different, there is a direct relationship between both formulations. Consider the set of signed integers $\{-n,\dots,-1,1,\dots,n\}$, and the permutation $\Gamma = ({-1} \ 1)({-2} \ 2)\cdots({-n} \ n)$. Then, going from chromosome to adjacency lists and back is achieved with a right multiplication by $\Gamma$~\citep{feijao2013extending} (with permutations acting on the left). 
For instance, recall that the circular signed chromosome shown in Figure~\ref{fig:oriented.genome}
is modelled by $\pi_c = (2\ {-4}\ 1 \ {-3}\ 6\ 5)(-5\ {-6}\ 3\ {-1}\ 4\ {-2})$, where each cycle
represents a strand of a chromosome, and signs represent gene orientation. 
Multiplying by $\Gamma$ on the right we get the adjacency representation $\pi_a$:
\begin{align*}
\pi_c\Gamma 
 &= (2\ {-4}\ 1 \ {-3}\ 6\ 5)(-5\ {-6}\ 3\ {-1}\ 4\ {-2}) ({-1} \ 1)({-2} \ 2)\cdots({-6} \ 6)\\
 &=  ({-2}\ {-4})(4\ 1)({-1}\ {-3})(3\ 6)({-6}\ 5)({-5}\ 2)\\
 &= \pi_a
\end{align*}
where negative and (omitted) positive signs represent the head and tail of each gene, respectively.

Since $\Gamma = \Gamma^{-1}$, this transformation works in both directions. It is also possible to apply this transformation for multi-chromosomal genomes, with both linear and circular genomes.

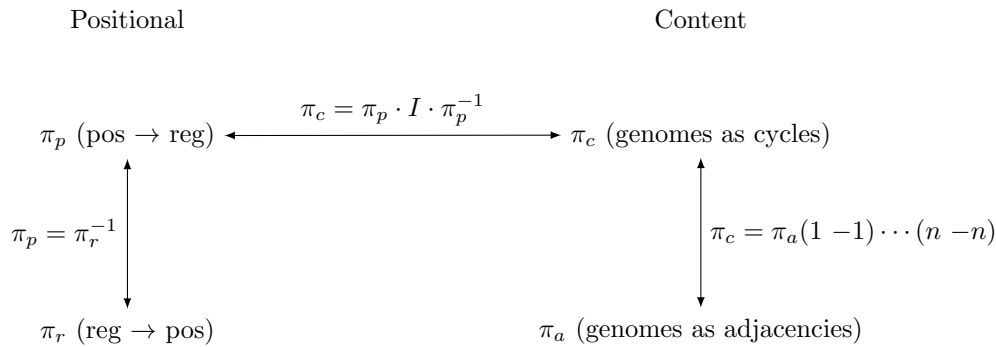
\begin{figure}[tph]  
\begin{center}
\begin{tikzpicture}[node distance=1cm]
\node (posreg) at (0,0) {$\pi_p$ (pos $\to$ reg)};
\node[below=of posreg, yshift=-1cm] (regpos) {$\pi_r$ (reg $\to$ pos)};
\node[above=of posreg] (pos) {Positional};

\node[right=of pos, xshift=5cm] (cont) {Content};
\node[below=of cont] (cycle) {$\pi_c$ (genomes as cycles)};
\node[below=of cycle, yshift=-1cm] (adj) {$\pi_a$ (genomes as adjacencies)};

\draw[>=latex,<->] (posreg)-- node[left]{$\pi_p = \pi_r^{-1}$} (regpos);
\draw[>=latex,<->] (cycle)-- node[right]{$\pi_c = \pi_a(1 \ {-1})\cdots(n \ {-n})$} (adj);
\draw[>=latex,<->] (posreg)-- node[above]{$\pi_c = \pi_p \cdot I \cdot \pi_p^{-1}$} (cycle);
\end{tikzpicture}
\end{center}
\caption{The relationship between permutations arising from different genome representations. $I = (1 \ 2 \ \dots \ n)$ for the unsigned case, or $I = (1 \ 2 \ \dots\ \ n)(-n \ {-n}+1 \ \dots \ {-1})$ for signed. Note that from $\pi_p$ to $\pi_c$ the translation is unique, while the opposite case has multiple possibilities (because there is choice in the labelling of positions). Also, the transformation from $\pi_c$ to $\pi_a$ is only possible --- or even the modelling of genomes as adjacencies in general --- for the signed (oriented) case.  All products here assume permutations act on the left.
}
\end{figure}

\subsection{Combined approaches}

We argue that some approaches in the genome rearrangement literature can, in fact, be interpreted as a combination of  the positional and the content paradigms. These approaches are based on interpreting the cycle graph of a permutation with an algebraic approach.

The \emph{cycle graph} of a permutation $\pi$, proposed by~\citet{Bafna1998}, is
a widely used graph where many results on genome rearrangement problems have
been obtained. \citet{Labarre2013}, extending results from a previous paper~\citep{Doignon2007}, 
proposed a framework that takes a
permutation $\pi$, representing a genome in the position paradigm, and
applies a transformation $f$, that is an algebraic representation of
the cycle graph of $\pi$. The transformation $f(\cdot)$ in question is given by
\[f(\pi) = (0 \ 1 \ \dots \ n)(\pi_n \ \pi_{n-1} \ \dots \ \pi_1 \ 0)\]
where $\pi_i = \pi(i)$, and a new element $0$ is added, to conform with the definition of the cycle graph, where
this element is also included. 
Clearly $f(\pi)$ is always an even permutation, and the authors show that the number of cycles of $f(\pi)$ is 
the same as the cycle graph of $\pi$, which allows them to find new rearrangement distance results  
based on the number of cycles and on decompositions of $f(\pi)$.

In the content paradigm as defined by \citet{meidanis2000alternative}, if we want to transform a genome
$\pi$ into another genome $\sigma$, we need operations $\rho_1,\rho_2,\dots,\rho_k$ such that
\[
\rho_k\cdots\rho_2\rho_1\pi = \sigma \implies \rho_k\cdots\rho_2\rho_1 = \sigma\pi^{-1}
\]
which means that decomposing of the permutation $\sigma\pi^{-1}$ is a way to obtain the rearrangement events, and the number of 
cycles of $\sigma\pi^{-1}$ is related to many rearrangement distances (e.g., \citep{Dias2001,feijao2013extending,bhatia2014algebraic}).
Now, if we complete the genomes with a $0$ element, we have that $\pi = (0 \ \pi_1 \ \pi_2 \dots \pi_n)$ and let the target genome $\sigma$ be the genome where all the elements are ``sorted'', that is, $\sigma = (0 \ 1 \ 2 \ \dots \ n)$. Therefore, the permutation $\sigma\pi^{-1}$ is exactly $f(\pi)$ as defined by Labarre. In that sense, we argue that the transformation proposed by Labarre is in fact acting as a translation between the position and content paradigms. 

\section{Summary and conclusions} 
\label{sec:summary}
In this paper, we have discussed three different models of a genome that are dominant in the genome rearrangement literature. The first of these views a chromosome as a map between a set of positions and a set of regions. The latter two model a genome as a map from a set of regions to itself.

In the position representation of a genome,  a rearrangement operator can be thought of as a map from the set of positions to itself. The positions are fixed, as against gene regions which `move' when a rearrangement operator acts on a genome. Thus a rearrangement operator modeled as a map from positions to positions can have the same form, {irrespective of the genome it is acting on}. This can simplify the problem of genome rearrangement in some cases, for example if we want to apply a fixed operator such as reversal operator, to a genome.

When a genome is viewed as a map from the set of regions to itself, genome rearrangement operators are also interpreted as maps from the set of regions to itself. This view of a genome has offered great modeling and algorithmic simplicity. For example, contrast the algorithms for sorting via the double cut and join operation \citet{yancopoulos2005efficient,bergeron2006unifying,feijao2013extending,bhatia2014algebraic} against the sorting by reversals algorithm in \citet{hannenhalli1999transforming}. A disadvantage of this approach is that the action of an operator depends on the structure of the genome on which it is acting. If we wish to fix the operation on the genome, then the set of operations available changes at each step as the genome is being sorted.

\end{document}